\documentclass[a4paper, 12pt]{article}
\usepackage{amsfonts,amsmath}

\begin{document}
\title{\noindent \textbf{The explicit secular equation for surface 
acoustic waves \\ in mo\-no\-cli\-nic elastic crystals}}

\author{Michel Destrade	\hspace*{\fill}\\
}
\date{2001}

\maketitle

\bigskip

%
\begin{abstract}
{\footnotesize
The secular equation for surface acoustic waves propagating 
on a monoclinic elastic half-space is  derived in a direct 
manner, using the method of first integrals.
Although the motion is at first 
assumed to correspond to generalized plane
strain, the analysis shows that only two components of the mechanical
displacement and of the tractions on planes parallel to the free 
surface are nonzero.
Using the Stroh formalism, a system of two second order differential 
equations is found for the remaining tractions.
The secular equation is then obtained as a quartic for the squared 
wave speed.
This explicit equation is consistent with that found in the 
orthorhombic case.
The speed of subsonic surface waves is then computed for twelve
specific monoclinic crystals.
}
\end{abstract}

\newpage


\section{INTRODUCTION}
{\normalsize
The modern theory of surface acoustic waves in anisotropic media 
owes most of its results to the pioneering works of A.~N.~Stroh.
Although his two seminal articles \cite{Stro58, Stro62} went largely 
unnoticed for a long time, their theoretical implications were 
far reaching, as many came to realize since their publication.
Among others,  Currie \cite{Curr74}, Barnett and Lothe \cite{BaLo74}, 
Chadwick and Smith \cite{ChSm77}, 
were able to use his `sextic formalism' to address many problems,
such as the existence of a single real secular equation for the wave 
speed, the existence of a limiting velocity (the smallest velocity
of body wave solutions) which defines `subsonic' and `supersonic'
ranges for the speeds, or numerical schemes to compute
the polarization vectors and the speed of the surface wave.
A comprehensive review of these topics can be found in a 
textbook by Ting \cite{Ting96}.

However precise numerical procedures might be, 
there is still progress to be made in the search for secular equations
in analytic form.
So far, explicit expressions have remained few.
The secular equation for surface waves in orthorhombic crystals was
established by Sveklo \cite{Svek48} as early as 1948  and later, 
Royer and Dieulesaint \cite{RoDi84} proved that it could account
for sixteen different crystal configurations, such as tetragonal,
hexagonal, or cubic.
For monoclinic media, Chadwick and Wilson \cite{ChWi92} devised 
a procedure to derive the secular equation, which is given as 
`explicit, [...] \textit{apart} from the solution of [a] bicubic 
equation.'
The object of this paper is to derive \textit{one} expression 
for the secular equation which is \textit{fully explicit}, 
when the surface wave propagates in monoclinic crystals.

A classical approach to the problem of surface waves in anisotropic
crystals is to consider that a wave propagates with speed $v$ in the
direction $x_1$ of a material axis (on the free plane surface)
of the material, and is attenuated along another material axis $x_2$,
orthogonal to the free surface, so that the mechanical displacement
$\mathbf{u}$ is written as $\mathbf{u}=\mathbf{u}(x_1 + p x_2 - vt)$,
where $p$ is unknown.
Then, assuming a complex exponential form for the displacement,
the equations of motion are written in the absence of body forces
and solved for $p$.
Finally, the boundary conditions yield the secular equation for $v$.
The principal mathematical difficulty arising from this procedure is
that the equations of motion yield a sextic (generalized
plane strain) or a quartic (plane strain) for $p$ which in general are
impractical to solve analytically, or even, as a numerical scheme 
suggests in the sextic case, are actually insoluble analytically 
(in the sense of Galois) \cite{Head79}.

In 1994,  Mozhaev \cite{Mozh94} proposed `some new ideas in the 
theory of surface acoustic waves.'
He introduced a novel method based on first integrals \cite{Yu83} 
of the displacement components, which bypasses the sextic 
(or quartic) equation for $p$ and yields directly  
the secular equation.
He successfully applied this method to the case of orthorhombic
materials.
In the present paper, generalized plane strain surface waves in a 
monoclinic crystal with plane of symmetry at $x_3=0$ are examined.
The method of first integrals is adapted in order to be applied to 
the tractions components on the planes $x_3=$const., rather than to 
the displacement components.
This switch  presents several advantages.
First, the equations of motion, the boundary conditions, 
and eventually the secular equation itself, 
are expressed directly in terms of the usual elastic stiffnesses.
Second, it makes it apparent that one of the traction components is 
zero and thus that, in this paper's context, generalized plane strain 
leads to plane stress.
Third, the boundary conditions are written in a direct and natural
manner, because they correspond to the vanishing of the tractions on
the free surface and at infinite distance from this surface.
Finally, this procedure can easily accommodate an internal constraint,
such as incompressibility \cite{NaSo97,SoNa99} 
(the secular equation for surface waves in
incompressible monoclinic linearly elastic materials is obtained 
elsewhere).

The plan of the paper is the following.
After a brief review of the basic equations describing motion in 
linearly elastic monoclinic materials (Section II),  
the equations of motion are written down in Section III for a surface 
acoustic wave with three displacement components which depend on two 
coordinates, that in the direction of propagation and that in the 
direction normal to the free surface (generalized plane strain).
Then in  Section IV, it is seen that one of the traction components
is identically zero (plane stress), and that consequently, so is 
one of the displacement components (plane strain).
For the remaining two traction components, coupled equations of motion
and the boundary conditions are derived in Section V.
Finally in Section VI, the method of first integrals is applied and
the secular equation for acoustic surface waves in monoclinic elastic 
materials is derived explicitly.
As a check,  the  subcase of orthorhombic materials is treated, 
and numerical results obtained by Chadwick and Wilson \cite{ChWi92}
for some monoclinic materials are recovered.

\section{PRELIMINARIES}

First,  the governing equations for a monoclinic elastic material 
are recalled.
The material axes of the media are denoted
by $x_1$, $x_2$, and $x_3$, and the plane $x_3=0$ is assumed to be
a plane of material symmetry.
For such a material, the relationship between the nominal stress 
$\mbox{\boldmath $\sigma$}$ and  the strain 
$\mbox{\boldmath $\epsilon$}$ is given by  \cite{Love27} 
\begin{equation} \label{StressStrainGeneral}
\begin{bmatrix} 
   \sigma_{11} \\
   \sigma_{22} \\
   \sigma_{33} \\
   \sigma_{23} \\
   \sigma_{31} \\
   \sigma_{12} \end{bmatrix}
=
\begin{bmatrix}
  c_{11} & c_{12} & c_{13} &    0   &   0   &  c_{16} \\
  c_{12} & c_{22} & c_{23} &    0   &   0   &  c_{26} \\
  c_{13} & c_{23} & c_{33} &    0   &   0   &  c_{36} \\
    0    &   0    &   0    & c_{44} & c_{45}&   0     \\
    0    &   0    &   0    & c_{45} & c_{55}&   0     \\
  c_{16} & c_{26} & c_{36} &    0   &   0   &  c_{66} \end{bmatrix}
\begin{bmatrix} 
   \epsilon_{11} \\
   \epsilon_{22} \\
   \epsilon_{33} \\
   2\epsilon_{23} \\
   2\epsilon_{31} \\
   2\epsilon_{12} \end{bmatrix},
\end{equation}
where  $c$'s denote the elastic stiffnesses, and the strain components 
$\epsilon$'s 
are related to the displacement components $u_1$, $u_2$, $u_3$ through
\begin{equation}
2 \epsilon_{ij}=(u_{i,j}+u_{j,i}) \quad (i,j=1,2,3).
\end{equation}
The equations of motion, written in the absence of body forces,
are
\begin{equation} \label{EqnMotnGeneral}
\sigma_{ij,j}= \rho u_{i,tt} \quad (i=1,2,3),
\end{equation}
where $\rho$ is the mass density of the material, 
and the comma denotes differentiation.

Finally, the $6\times 6$ matrix $\mathbf{c}$ given in 
Eq.~\eqref{StressStrainGeneral} must be positive definite in order for
the strain-energy function density to be  positive.

\section{SURFACE WAVES}

Now the propagation of a  surface wave on a semi-infinite 
body of monoclinic media is modeled.
In the same manner as Mozhaev \cite{Mozh94}, the amplitude of the 
associated displacement is assumed to be varying sinusoidally with
time in the direction of propagation $x_1$, whilst its variation
in the direction $x_2$, orthogonal to the free surface, is not stated
explicitly.
Thus, calling $v$ the speed of the wave, and $k$ the associated
wave number, the displacement components are written in the form 
\begin{equation}
u_j(x_1,x_2,x_3,t)
=U_j(x_2) e^{ik(x_1 -vt)} \quad (j=1,2,3),
\end{equation}
where the $U$'s depend on $x_2$ only.
For these waves, the planes of constant phase are orthogonal to the
$x_1$-axis, and the planes of constant amplitude are orthogonal to the
$x_2$-axis.

The stress-strain relations \eqref{StressStrainGeneral} reduce to 
\begin{equation}\label{StressStrain}
\begin{array}{l}
t_{11}=i c_{11}U_1 +c_{12}U_2' +c_{16}(U_1' + i U_2),
\\
t_{22}=i c_{12}U_1 +c_{22}U_2' +c_{26}(U_1' + i U_2),
\\
t_{33}=i c_{13}U_1 +c_{23}U_2' +c_{36}(U_1' + i U_2),
\\
t_{32}=c_{44} U_3' +i c_{45}U_3, \\  
t_{13}=c_{45} U_3' +i c_{55} U_3, \\
t_{12}=i c_{16}U_1 +c_{26}U_2' +c_{66}(U_1' + i U_2), 
\end{array}
\end{equation}
where the prime denotes differentiation with respect to $k x_2$,
and the $t$'s are defined by 
\begin{equation} 
\sigma_{ij}(x_1,x_2,x_3,t)= k t_{ij}(x_2)e^{ik(x_1 -vt)}
\quad (i,j=1,2,3).
\end{equation}

The boundary conditions of the problem 
(surface $x_2=0$ free of tractions, 
vanishing displacement as $x_2$ tends to infinity) are
\begin{equation} \label{BC1}
t_{i2}(0)=0, \quad U_i(\infty)=0 
\quad (i=1,2,3).
\end{equation}

Finally, the equations of motion \eqref{EqnMotnGeneral}  reduce to 
\begin{equation} \label{EqnMotn}
\begin{array}{l}
i t_{11}+ t_{12}'= - \rho v^2 U_1, \:
i t_{12}+ t_{22}'= - \rho v^2 U_2, \:
i t_{13}+ t_{32}'= - \rho v^2 U_3.
\end{array}
\end{equation}

At this point,a sextic formalism could be developed for the three
displacement components $U_1$, $U_2$, $U_3$, and the three traction
components $t_{12}$, $t_{22}$, $t_{32}$.
However, it turns out that one of these traction components is 
identically zero, as is now proved.

\section{PLANE STRESS}

It is known (see the Appendix of Stroh's 1962 paper \cite{Stro62},
and also Ting's book \cite{Ting96}, p.66) that for a two-dimensional
deformation of a monoclinic crystal with axis of symmetry at $x_3=0$,
the displacements $u_1$ and $u_2$ are decoupled from $u_3$.
Taking $u_3=0$ for surface waves, it follows 
from the stress-strain relationships \eqref{StressStrain} that 
$t_{13}=t_{32}=0$.
Here, an alternative proof of this result is presented.

Using Eqs.~\eqref{StressStrain}$_4$, \eqref{EqnMotn}$_3$, and
 \eqref{StressStrain}$_5$, two first order differential equations for 
$t_{32}$ and $U_3$ are found as
\begin{equation} \label{t3-U3}
t_{32}=i c_{45} U_3 + c_{44} U_3', \quad
t_{32}' = (c_{55} - \rho v^2) U_3 - i c_{45} U_3'.
\end{equation}
These equations may be inverted to give $U_3$ and $U_3'$ as
\begin{equation} \label{U3-U3'}
\begin{array}{l}
(c_{44} c_{55} - c_{45}^2 - c_{44} \rho v^2) U_3
 = i c_{45} t_{32} + c_{44} t_{32}', \\
(c_{44} c_{55} - c_{45}^2 - c_{44} \rho v^2) U_3'
 =  (c_{55} - \rho v^2) t_{32} - i c_{45} t_{32}'.
\end{array}
\end{equation}

Differentiation of \eqref{U3-U3'}$_1$ and comparison with 
\eqref{U3-U3'}$_2$ yields the following second order differential 
equation for $t_{32}$,
\begin{equation} \label{t32}
c_{44} t_{32}'' + 2i c_{45} t_{32}' - (c_{55} - \rho v^2) t_{32} =0.
\end{equation}
The boundary conditions \eqref{BC1} and Eq.~\eqref{StressStrain}$_4$
imply that the stress component $t_{32}$ must satisfy 
$t_{32}(0)= t_{32}(\infty)=0$.
The only solution of this boundary value problem for the differential 
equation \eqref{t32} is the trivial one.
Consequently, 
\begin{equation}
t_{32}(x_2)=0 \quad \text{for all } x_2,
\end{equation}
and so it is proved that, as far as the propagation of surface
acoustic waves in monoclinic crystals with plane of symetry at
$x_3=0$ is concerned, generalized plane strain leads to plane stress.

It is also worth noting that by Eq.~\eqref{U3-U3'}$_1$, plane stress 
leads in turn to plane strain which, as an assumption, was not 
needed a priori.
This result was obtained by Stroh \cite{Stro62} in a different manner:
 `[when] there is a reflection plane normal to the $x_3$ axis, 
[...] there is no coupling of the displacement $u_3$ with $u_1$ 
and $u_2$;
any two dimensional problem reduces to one of plane strain 
($u_3=0$) and one of anti-plane strain ($u_1=u_2=0$).'

Now the equations of motion can be written for the remaining 
displacements and traction components.

\section{EQUATIONS OF MOTION}

Here, the equations of motion are derived, first as a system of four
first order differential equations for the nonzero components of 
mechanical displacement and tractions, and then as a system of two
second order differential equations for the tractions.

The stress-strain relations \eqref{StressStrain} and the 
equations of motion \eqref{EqnMotn} lead to a system of differential
equations for the displacement components $U_1$, $U_2$, and for
the traction components $t_1$, $t_2$, defined by
\begin{equation} \label{t1t2}
t_1 = t_{12}, \quad t_2 = t_{22}.
\end{equation}
This system is as follows
\begin{equation} \label{Stroh}
\begin{bmatrix}
 \mathbf{u}' \\
 \mathbf{t}' \end{bmatrix}
=
\begin{bmatrix}
 i \mathbf{N_1} & \mathbf{N_2} \\
 -(\mathbf{N_3} + X \mathbf{1}) & i \mathbf{N_1}^{\mathrm{T}}
\end{bmatrix}
\begin{bmatrix}
 \mathbf{u} \\
 \mathbf{t} \end{bmatrix},
\end{equation}
where $\mathbf{u}= [ U_1, U_2  ]^{\mathrm{T}}$, 
$\mathbf{t}= [ t_1, t_2  ]^{\mathrm{T}}$, $X= \rho v^2$, 
and the $2 \times 2$ matrices $\mathbf{N_1}$, $\mathbf{N_2}$, 
and  $\mathbf{N_3}$ are submatrices of the fundamental elasticity
matrix $\mathbf{N}$, introduced by Ingebrigsten and Tonning
\cite{InTo69}.
Explicitly, $\mathbf{N_1}$, $\mathbf{N_2}$, $\mathbf{N_3}$ are
 given by \cite{Ting96}
\begin{equation}
\begin{array}{c}
-\mathbf{N_1} = 
   \begin{bmatrix}
     r_6 & 1 \\
     r_2 & 0
    \end{bmatrix},
 \quad
\mathbf{N_2} = 
\begin{bmatrix}
     s_{22} & -s_{26} \\
    -s_{26} &  s_{66} 
    \end{bmatrix}
 = \mathbf{N_2}^{\mathrm{T}}, 
 \quad
-\mathbf{N_3} = 
\begin{bmatrix}
    \eta &  0 \\
      0  &  0
    \end{bmatrix}
 = - \mathbf{N_3}^{\mathrm{T}}, 
\end{array}
\end{equation}
where the quantities $r_2$, $r_6$, $s_{22}$, $s_{26}$, $s_{66}$, 
and $\eta$ are given in terms of the elastic stiffnesses as
\begin{align} \label{Coefficients}
\Delta &= \begin{vmatrix}
     c_{22} & c_{26} \\
     c_{26} & c_{66}
    \end{vmatrix}
       = c_{22} c_{66} - c_{26}^2,
\nonumber \\ 
r_6 &=  \frac{1}{\Delta}(c_{22} c_{16} - c_{12}c_{26}) 
\nonumber  \\
r_2 &=  \frac{1}{\Delta}(c_{12} c_{66} - c_{16}c_{26}), \\
s_{ij} &=  \frac{1}{\Delta}c_{ij} \quad (i,j = 2,6),    
\nonumber  \\   
\eta &=  \frac{1}{\Delta}
    \begin{vmatrix}
     c_{11} & c_{12} & c_{16} \\  
     c_{12} & c_{22} & c_{26} \\  
     c_{16} & c_{26} & c_{66}
    \end{vmatrix}
    =   c_{11} 
     -  \frac{c_{66}c_{12}^2 + c_{22}c_{16}^2 - 2 c_{12}c_{16}c_{26}}
               {c_{22} c_{66} - c_{26}^2}.\nonumber  
\end{align}

Throughout the paper, it is assumed that the matrix 
$\mathbf{N_3} + X \mathbf{1}$ is not singular, 
which means that the surface wave propagates at a speed distinct from
that given by $\rho v^2 = \eta$.
This assumption made, the second vector line of the system 
\eqref{Stroh} yields
\begin{equation} \label{u-t-t'}
\mathbf{u} = i(\mathbf{N_3} + X \mathbf{1})^{-1}
 \mathbf{N_1}^{\mathrm{T}} \mathbf{t}
 - (\mathbf{N_3} + X \mathbf{1})^{-1}\mathbf{t}'.
\end{equation}
On the other hand, differentiation of the system \eqref{Stroh} leads to
\begin{equation}
\begin{bmatrix}
 \mathbf{u}'' \\
 \mathbf{t}'' \end{bmatrix}
=
\begin{bmatrix}
 - \mathbf{N_1} \mathbf{N_1}
   - \mathbf{N_2}(\mathbf{N_3} + X \mathbf{1}) 
 &i(\mathbf{N_1} \mathbf{N_2}+ \mathbf{N_2}\mathbf{N_1}^{\mathrm{T}})
 \\
 -i[(\mathbf{N_3} + X \mathbf{1})\mathbf{N_1} +
  \mathbf{N_1}^{\mathrm{T}}(\mathbf{N_3} + X \mathbf{1})] 
  & -(\mathbf{N_3} + X \mathbf{1}) \mathbf{N_2} - 
    \mathbf{N_1}^{\mathrm{T}}\mathbf{N_1}^{\mathrm{T}}
\end{bmatrix}
\begin{bmatrix}
 \mathbf{u} \\
 \mathbf{t} \end{bmatrix},
\end{equation}
Now the second vector line of this equation yields, using 
Eq.~\eqref{u-t-t'}, a system of two second order differential equations
for $\mathbf{t}$, written as
\begin{equation} \label{syst-t}
\widehat{\alpha}_{ik} t_k'' - i \widehat{\beta}_{ik} t_k'
 - \widehat{\gamma}_{ik} t_k = 0, 
\end{equation}
where the symmetric $2 \times 2$ matrices 
\mbox{\boldmath $\widehat{\alpha}$},
\mbox{\boldmath $\widehat{\beta}$}, 
and \mbox{\boldmath $\widehat{\gamma}$},
are given by
\begin{equation}
\begin{array}{c}
\mbox{\boldmath $\widehat{\alpha}$} 
= -(\mathbf{N_3} + X \mathbf{1})^{-1},
\quad
\mbox{\boldmath $\widehat{\beta}$} = 
  -\mathbf{N_1}(\mathbf{N_3} + X \mathbf{1})^{-1}- 
  (\mathbf{N_3} + X \mathbf{1})^{-1}\mathbf{N_1}^{\mathrm{T}}
\\
\mbox{\boldmath $\widehat{\gamma}$} =  \mathbf{N_2}-
 \mathbf{N_1}(\mathbf{N_3}+X\mathbf{1})^{-1}\mathbf{N_1}^{\mathrm{T}},
\end{array}
\end{equation}
or, explicitly, by their components,
\begin{align} \label{AlphaBetaGamma}
\vspace{10pt}
\widehat{\alpha}_{11}& = \frac{1}{\eta - X}, \quad
\widehat{\alpha}_{12} = 0, \quad
\widehat{\alpha}_{22} = - \frac{1}{X}, \nonumber \\ 
\widehat{\beta}_{11} &= - \frac{2 r_6}{\eta - X}, \quad
\widehat{\beta}_{12} = \frac{1}{X} - \frac{r_2}{\eta - X}, \quad 
\widehat{\beta}_{22} = 0, \\
\widehat{\gamma}_{11}& = s_{22} + \frac{r_6^2}{\eta - X}
 - \frac{1}{X},  \quad
\widehat{\gamma}_{12} =\frac{r_2 r_6}{\eta - X} - s_{26}, \quad 
\widehat{\gamma}_{22} = \frac{r_2^2}{\eta - X} + s_{66}. \nonumber
\end{align}

The system \eqref{syst-t} of second order differential equations for
the traction components is more convenient to work with than the 
corresponding system for the displacement components, 
because the  boundary conditions are simply written, 
using Eqs.~\eqref{BC1}, \eqref{StressStrain}, and \eqref{t1t2}, as
\begin{equation} \label{BC2}
t_{i}(0)=  t_i(\infty)= 0 
\quad (i=1,2).
\end{equation}
This claim is further justified in the next section, where the secular
equation is quickly derived.

\section{SECULAR EQUATION}

Now the method of first integrals is applied to the system  
\eqref{syst-t}.
Mozhaev \cite{Mozh94} defined the following inner product,
\begin{equation}
(f, \phi) =  \textstyle{\int} (f \overline{\phi} +\overline{f} \phi)
dx_2,
\end{equation}
and multiplying Eq.~\eqref{syst-t} by $i \overline{t_j}$ gives
\begin{equation} \label{syst-DEF}
\widehat{\alpha}_{ik} D_{kj} + \widehat{\beta}_{ik} E_{kj}
 + \widehat{\gamma}_{ik} F_{kj} = 0, 
\end{equation}
where the $2 \times 2$ matrices $\mathbf{D}$, $\mathbf{E}$,
$\mathbf{F}$, are defined by
\begin{equation} 
D_{kj} = (i t_k'', t_j), \quad
E_{kj} = (t_k', t_j), \quad
F_{kj} = (t_k, i t_j).
\end{equation}

By writing down $F_{kj}+F_{jk}$, it is easy to check that the matrix
$\mathbf{F}$ is antisymmetric.
Integrating directly $E_{kj}+E_{jk}$, and integrating $D_{kj}+D_{jk}$ 
by parts, and using the boundary conditions \eqref{BC2}, 
it is found that the matrices $\mathbf{E}$  and $\mathbf{D}$ are also
antisymmetric.
So $\mathbf{D}$, $\mathbf{E}$, and $\mathbf{F}$ may be written in the
form
\begin{equation}
\mathbf{D}= \begin{bmatrix} 0 & D \\ -D & 0 \end{bmatrix} \quad
\mathbf{E}= \begin{bmatrix} 0 & E \\ -E & 0 \end{bmatrix} \quad
\mathbf{F}= \begin{bmatrix} 0 & F \\ -F & 0 \end{bmatrix},
\end{equation}
and Eq.~\eqref{syst-DEF} yields the following system of three linearly
independent equations for the three unknowns $D$, $E$, $F$,
\begin{equation}
\begin{array}{l}
\widehat{\alpha}_{11}D +\widehat{\beta}_{11}E +\widehat{\gamma}_{11}F
 =0, \\
\widehat{\alpha}_{12}D +\widehat{\beta}_{12}E +\widehat{\gamma}_{12}F
 =0, \\
\widehat{\alpha}_{22}D +\widehat{\beta}_{22}E +\widehat{\gamma}_{22}F
 =0.
\end{array}
\end{equation}
This homogeneous linear algebraic system yields nontrivial solutions
for $D$, $E$, and $F$, only when its determinant is zero,
which, accounting for the fact that 
$\widehat{\alpha}_{12}=\widehat{\beta}_{22}=0$, is equivalent to
$\widehat{\beta}_{12}(\widehat{\alpha}_{11}\widehat{\gamma}_{22}
-\widehat{\alpha}_{22}\widehat{\gamma}_{11}) = 
-\widehat{\alpha}_{22}\widehat{\beta}_{11}\widehat{\gamma}_{12}$, 
or equivalently, using the expressions \eqref{AlphaBetaGamma} 
and multiplying by $X^3(\eta-X)^3$,
\begin{multline} \label{Secular}
[\eta - (1+r_2)X] 
  \{ (\eta -X)[(\eta -X)(s_{22}X -1) + r_6^2 X]
      + X^2[(\eta -X)s_{66} + r_2^2] \} \\
= 2 r_6 X^2 (\eta -X) [ (\eta -X)s_{26} - r_2 r_6].
\end{multline}
Hence \textit{the secular equation is obtained explicitly 
as the quartic \eqref{Secular} in $X=\rho v^2$}, 
with coefficients expressed in terms
of the elastic stiffnesses through Eqs.~\eqref{Coefficients}.

For consistency, the orthorhombic case, where 
$c_{16}=c_{26}=c_{45}=0$, is now considered.
In this case, the coefficients \eqref{Coefficients} reduce to 
\begin{equation} 
r_6=0, \quad r_2=\frac{c_{12}}{c_{22}}, \quad 
s_{22}=\frac{1}{c_{66}}, \quad s_{26}=0, \quad 
s_{66}=\frac{1}{c_{22}}, \quad \eta = c_{11}-\frac{c_{12}^2}{c_{22}},
\end{equation}
and the right hand-side of Eq.~\eqref{Secular} is zero,
while the left hand-side yields the equation
\begin{equation} \label{Ortho}
[\eta - (1+r_2)X] 
  \{ (\eta -X)^2(s_{22}X-1)
      + X^2[(\eta -X)s_{66} + r_2^2] \}
= 0.
\end{equation}
The nullity of the first factor in this equation corresponds to 
$\widehat{\beta}_{12}=0$.
Because for the orthorhombic case,
$\widehat{\alpha}_{12}=\widehat{\gamma}_{12}=\widehat{\beta}_{11}
=\widehat{\beta}_{22}=0$ also,
the equations of motion \eqref{syst-t} then decouple into 
\begin{equation} 
\widehat{\alpha}_{11} t_1'' + \widehat{\gamma}_{11} t_1 = 0, \quad 
\widehat{\alpha}_{22} t_2'' + \widehat{\gamma}_{22} t_2 = 0,
\end{equation}
whose solutions satisfying the boundary conditions \eqref{BC2} are the
trivial ones.
The nullity of the second factor in Eq.~\eqref{Ortho} corresponds to 
the well-studied \cite{Svek48, Ston63, Ting96} secular equation for 
surface waves in orthorhombic crystals,
\begin{equation} 
\frac{c_{22}}{c_{11}}
	(\frac{c_{11} c_{22} - c_{12}^2}{c_{22} c_{66}} 
		- \frac{\rho v^2}{c_{66}})^2
(1 - \frac{\rho v^2}{c_{66}}) - 
 (\frac{\rho v^2}{c_{66}})^2 (1 - \frac{\rho v^2}{c_{11}}) = 0.
\end{equation}

Finally, concrete examples are given (see Table 1).
In each considered case, the secular equation \eqref{Secular} has 
either 2 or 4 positive real roots, out of which only one corresponds 
to a subsonic wave.
The elimination of the other roots is made by comparison with the 
speed of a homogeneous body wave propagating in the direction of the
$x_1$ material axis.
For this body wave, the functions $U_i(x_2)$, $t_i(x_2)$, ($i=1,2$),
are constant, and the equations of motion imply that the determinant
of the $4 \times 4$ matrix in Eq.~\eqref{Stroh} is zero, condition from
which the body wave speed can be found.
Also, it is checked a posteriori that the value $X=\eta$ corresponds
to the supersonic range, and so that the matrix $N_3 + X \mathbf{1}$ 
is indeed invertible within the subsonic range.
For instance, for tin fluoride,
$\eta$ is of the order of $3 \times 10^7$, 
the secular equation \eqref{Secular}
has the roots 1339, 2350, 2513, and 3403, 
and the slowest body wave in the $x_1$ direction travels at 
1504 m s$^{-1}$; 
hence a subsonic surface wave travels in tin fluoride 
at 1339  m s$^{-1}$.

\begin{center}
Table 1.
\textit{Values of the relevant elastic stiffnesses} (GPa), 
\textit{density} (kg   m$^{-3}$),
\textit{and surface wave speed} (m  s$^{-1}$) 
\textit{for 12 monoclinic crystals.}

\noindent
{\small
\begin{tabular}{l c c c c c c c c}
\hline
\rule[-3mm]{0mm}{8mm} 
material & $c_{11}$  & $c_{22}$     & $c_{12}$    & $c_{16}$
 & $c_{26}$      & $c_{66}$    & $\rho$     & $v$
\\
\hline
aegirite-augite & 216 & 156 & 66 & 19 & 25 & 46.5 & 3420 & 3382
\\
augite & 218 & 182 & 72 & 25 & 20 & 51.1 & 3320 & 3615
\\
diallage & 211 & 154 & 37 & 12 & 15 & 62.2 & 3300 & 4000
\\
diopside & 238 & 204 & 88 & -34 & -19 & 58.8 & 3310 & 3799
\\
diphenyl & 14.6 & 5.95 & 2.88 & 2.02 & 0.40 & 2.26 & 1114 & 1276
\\
epidote & 202 & 212 & 45 & -14.3 & 0 & 43.2 & 3400 & 3409
\\
gypsum & 50.2 & 94.5 & 28.2 & -7.5 & -11.0 & 32.4 & 2310 & 3011
\\
hornblende & 192 & 116 & 61 & 10 & 4 & 31.8 & 3120 & 3049
\\
microcline & 122 & 66 & 26 & -13 & -3 & 23.8 & 2561 & 2816
\\
oligoclase & 124 & 81 & 54 & -7 & 16 & 27.4 & 2638 & 2413
\\
tartaric acid & 46.5 & 93 & 36.7 & -0.4 & -12.0 & 8.20 & 1760 & 1756
\\
tin fluoride & 33.6 & 47.9 & 5.3 & 6.5 & -5.1 & 12.9 & 4875 & 1339
\\
 \hline
\end{tabular}
}
\end{center}

Barnett, Chadwick, and Lothe \cite{BaCL91},
and  Chadwick and Willson \cite{ChWi92} considered surface waves
propagating in monoclinic materials, 
and computed the surface wave speed $v$ in two steps,
first by solving numerically a bicubic,
then by substituting the result into another equation of which $v$ is
the only zero.
These authors studied surface wave propagation for every value of the
angle $\alpha$ between the reference plane and the plane of material
symmetry.
Numerical values for $v$ are only given in the cases of 
aegirite-augite, diallage, gypsum, and microcline, 
and at $\alpha=0$, these results are in agreement with those
presented in Table 1.
Sources of experimental data and extensive discussions on
limiting speeds, existence of secluded supersonic surface waves,
rotation of the reference plane with respect to the plane of 
material symmetry, etc., can  be found in these articles 
and in references therein.

\section{CONCLUDING REMARKS}

Surface wave motion in monoclinic crystals with plane of symmetry at
$x_3=0$ turned out to correspond to plane strain and plane stress 
motion (Section IV).
Thanks to this, the equations of of motion yielded a system of only two
differential equations for the tractions (Section V).
Once the method of first integrals was applied, a homogeneous 
system of 3 linearly independent equations for three unknowns
was obtained (Section VI).
Had the motion not corresponded to plane stress, then the same 
procedure would have given a system of 18 equations for 18 unknowns, 
when the equations of motion are written for the
displacement components  \cite{Mozh94},
or a system of 9 equations for 9 unknowns, 
when the equations of motion are written for the
traction components as in the present paper.
However these equations are not linearly independent, and the secular 
equation cannot be obtained in this manner.
Hence, it ought to be stressed again that the method presented in the 
paper is not a general method for a surface wave traveling in 
arbitrary direction in an anisotropic crystal,
but was limited to the study of a surface wave propagating in the 
$x_1$-direction of a monoclinic crystal with plane of symmetry at
$x_3=0$, with attenuation in the $x_2$-direction.

Nevertheless, some plane strain problems remain open and it is hoped 
that the method exposed in this paper might help solve them 
analytically.
Also, beyond mathematical satisfaction, the derivation of an explicit
secular equation provides a basis for a possible nonlinear 
perturbative analysis.


\end{document}